%% file: Main.tex
\documentclass[conference]{IEEEtran}
\IEEEoverridecommandlockouts

\usepackage{newtxtext, newtxmath} 

\usepackage{microtype}

\usepackage{booktabs}
\usepackage{cite}
\usepackage{amsmath}
\usepackage{graphicx}
\usepackage{textcomp}
\usepackage{xcolor}

\usepackage{flushend}
\usepackage[detect-all]{siunitx}
\usepackage{url}
\usepackage{booktabs}
\usepackage{dblfloatfix}
\usepackage{cuted}
\usepackage{tablefootnote}
\usepackage{eso-pic}

\usepackage[font=footnotesize]{caption}
\begin{document}

\title{Environmental Assessment of GaN Current-Source and Voltage-Source AC-AC Variable Speed Drives}

\author{Luc Imperiali,~\IEEEmembership{Student Member,~IEEE,}

        and~Jonas Huber,~\IEEEmembership{Senior Member,~IEEE}
\thanks{L. Imperiali is with the Advanced Mechatronic Systems Group, ETH Zurich, Switzerland, e-mail: imperiali@ams.ee.ethz.ch.}
\thanks{J. Huber is with the Power Electronics and Drive Systems Laboratory, ETH Zurich, Switzerland.}
}

\maketitle

\AddToShipoutPictureBG*{%
  \AtPageLowerLeft{%
    \put(\LenToUnit{0.5\paperwidth},\LenToUnit{0.8cm}){%
      \makebox[0pt][c]{%
        \begin{minipage}{0.95\textwidth}
          \centering\footnotesize
          This work has been submitted to the IEEE for possible publication.\\
          Copyright may be transferred without notice, after which this version may no longer be accessible.
        \end{minipage}%
      }%
    }%
  }%
}

\begin{abstract}
Three-phase ac-ac converters for variable-speed drives can be realized conventionally as voltage-source converters (VSCs) or, leveraging emerging monolithic bidirectional GaN power transistors (M-BDSs), as current-source converters (CSCs). 
This paper presents a comparative life-cycle assessment of a realized VSC and a CSC demonstrators including grid-side and motor-side EMI filters, and with identical specifications (1.4\,kW rated power, 200\,V line-to-line rms grid and motor voltage). 
The embodied carbon footprints are similar (\qty{44.8}{kg\,CO_2eq} for the VSC and \qty{47.4}{kg\,CO_2eq} for the CSC) and dominated by auxiliary and control electronics, and the printed circuit boards. 
Over \qty{10}{years} of continuous operation under the NEMA Power Index mission profile for variable-load centrifugal systems (pumps, fans), the CSC’s higher part-load efficiency translates into a clearly lower life-cycle carbon footprint of around 900\,kg\,CO\textsubscript{2}eq compared to the VSC’s 1130\,kg\,CO\textsubscript{2}eq when operating with a carbon-intense electricity mix (440\,g\,CO\textsubscript{2}eq/kWh). However, this difference almost vanishes when considering a future renewable electricity mix (25\,g\,CO\textsubscript{2}eq/kWh).

\end{abstract}


\input{Text}

\section{Acknowledgment}
The authors would like to thank Dr. Neha Nain from Infineon Technology for sharing the design details of the two demonstrator systems.
Further the authors would also like to thank the \textit{Swiss Federal Office of Energy} for the financial support of this research. The Advanced Mechatronic Systems Group at ETH Zurich is generously supported by the \textit{Else und Friedrich Hugel Fonds} via the \textit{ETH Zurich Foundation}, for which the authors are most grateful.


\bibliographystyle{IEEEtran}
\bibliography{bibliography}
	
\end{document}

%% file: Text.tex
\section{Introduction}
Electric motor-driven systems account for nearly half of global electricity use \cite{ieaEnergyEfficiencyPolicy2011}, and in manufacturing industries that share can reach about 80\% \cite{malinowskiDepartmentEnergyMotor2023}.
In such industrial environments, variable-load machines such as pumps, fans, and compressors account for a large part of that demand, and many are still driven by grid-fed motors running at a fixed speed, with valves or dampers setting the mechanical output, thus wasting significant amounts of energy  \cite{malinowskiDepartmentEnergyMotor2023}.
In such variable-load centrifugal systems, the torque typically rises with the square of speed, thus the motor speed is a direct lever for reducing the power consumption.
A variable-speed drive (VSD) placed between the grid and the machine can adjust the motor speed, and thus the power output of the motor, to the demand, and the energy use falls sharply.

\begin{figure*}[!]
    \centering
    \includegraphics{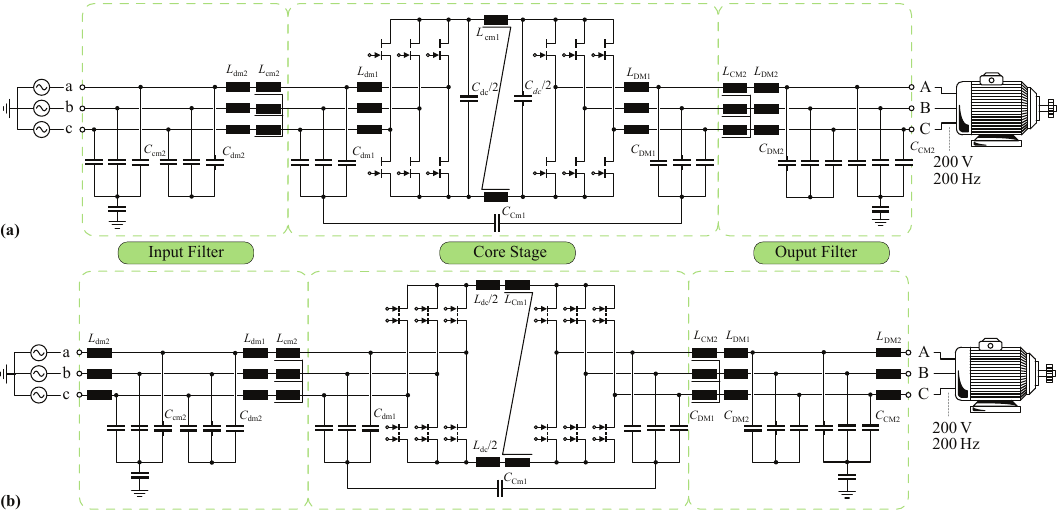}
    \captionof{figure}{Three-phase ac-ac variable-speed drives (VSDs), consisting of a core stage and grid- and motor-side EMI filters.
    \textbf{(a)}~Voltage-source converter (VSC) with conventional GaN power transistors and
    \textbf{(b)}~current-source converter (CSC) with monolithic bidirectional GaN power transistors.
    Figure adapted from \cite{Neha_CSC_VSC}.}
    \label{fig:schematics}
\end{figure*}

The typical realization of a grid-connected VSD is a two-level voltage-source converter (VSC), i.e., a back-to-back connection of a three-phase voltage-source rectifier and a three-phase inverter sharing a common dc-link capacitor.
Modern voltage-source converters (VSCs) increasingly employ wide-bandgap semiconductors such as GaN power transistors, which enable higher switching frequencies and, consequently, smaller passive components.
However, the resulting fast voltage transitions (high $dv/dt$) at the inverter output can increase motor bearing currents, cause reflections on the motor cables and accelerate motor insulation ageing, and may require shielded motor cables to avoid EMI issues \cite{Neha_CSC_VSC}. To mitigate these effects, motor-side and grid-side EMI filters are typically employed as shown in \textbf{Fig.~\ref{fig:schematics}a}.
An alternative topology for a grid-connected VSD is the current-source converter (CSC), which consists of a three-phase current-source rectifier and a three-phase current-source inverter sharing a common dc-link inductor, see \textbf{Fig.~\ref{fig:schematics}b}. Unlike a VSC, the switching stage of the CSC provides a switched \textit{current} at its phase terminals, i.e., requires ac-side capacitors that absorb the switched current and thereby inherently ensure smooth output voltages. Nevertheless, grid-side and motor-side EMI filters are still required to meet specific emission limits.

Again, unlike a VSC, the voltage across a switch of a CSC is not a dc voltage but defined by the ac line-to-line voltage and, hence, can have either polarity. Consequently, each switch must be capable of blocking both positive and negative voltages. Traditionally, this capability was achieved by connecting two unipolar transistors in anti-series, doubling the on-state resistance compared to a single transistor. Achieving the same total on-state resistance thus requires paralleling of a second anti-series connection of two transistors, i.e., a four-times higher chip area overall. 

This structural drawback of the CSC has recently been overcome by the advent of monolithic bidirectional GaN power transistors (M-BDSs). First-generation \qty{600}{V}, \qty{140}{\mohm} GaN M-BDSs use a common drift region for blocking both voltage polarities, avoiding the area penalty associated with conventional back-to-back transistor arrangements \cite{Bisi_GaN_BDS,Infineon_HV_GaN_BDS}. This development makes it possible to implement the CSC with a semiconductor die area comparable to that of the VSC, enabling a meaningful comparison of the two converter topologies.
Such a comparison based on detailed experimental results has been described in \cite{Neha_CSC_VSC} for two respective \qty{1.4}{kW} demonstrators with identical specifications (see \textbf{Tab.~\ref{tab:key_specifications}}), in particular meeting 
the same conducted EMI limits on the grid and on the motor (CISPR~11 Class~A and IEC~TS~62578~C2),
evaluating power densities, efficiency profiles, and EMI compliance. 

Clearly, VSDs enable massive gains in energy efficiency of motor-driven variable-load centrifugal systems. However, complementing traditional key performance indicators (KPIs) like efficiency and power density, there is growing interest from industry \cite{musilHowLifeCycle2023a, camposCO2FootprintMedium2024, guillemetOpenSourceLifeCycle2026}, regulatory bodies \cite{RegulationEU20242024}, and academia \cite{nordelofScalableLifeCycle2019,radwanLifeCycleAssessment2025,fangParametricLCAModel2025,baudaisLifeCycleAssessment2023,ningwangEcodesignMagneticComponents, imperialiMultiobjectiveMinimizationLifecycle2024, huMethodSustainableAnalysis2025} in life-cycle assessment (LCA) and ecodesign of power converters.

The VSC and CSC demonstrators from \cite{Neha_CSC_VSC} are built to meet the same specifications, yet employ fundamentally different topologies. It is therefore necessary to evaluate how the two systems compare with respect to their embodied carbon footprints as well as with respect to the life-cycle carbon emissions under a relevant mission profile.

Therefore, this paper extends the hardware comparison of \cite{Neha_CSC_VSC} accordingly, following a similar approach as in a recent study regarding two different realization options for ultralightweight dc--dc converters \cite{Imperiali_dc_dc}. In the following, \textbf{Section~\ref{sec:hwimplementation}} describes and explains the relevant aspects of the analyzed hardware, \textbf{Section~\ref{sec:embodied}} estimates the embodied carbon footprints, including a variant of the CSC with reduced gate drive realization effort. Finally, \textbf{Section~\ref{sec:use_phase}} discusses use-phase carbon emissions before \textbf{Section~\ref{sec:conclusion}} concludes the paper.

\section{Hardware Implementation}\label{sec:hwimplementation}

\begin{figure}[h!]
    \centering
    \includegraphics[width=\columnwidth]{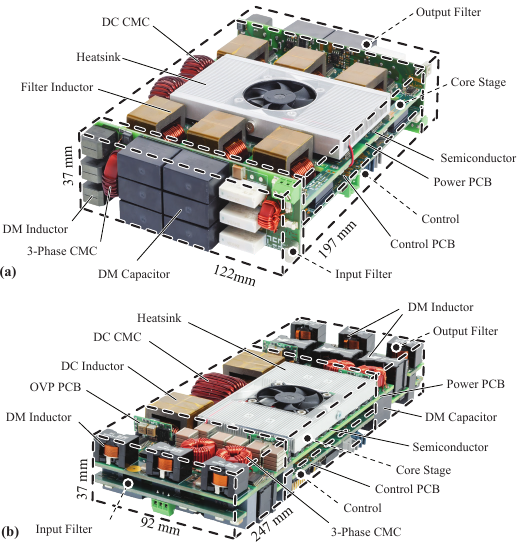}
    \caption{Realized \qty{1.4}{kW} demonstrators of the VSDs from \textbf{Fig.~\ref{fig:schematics}} with realization details given in \textbf{Tab.~\ref{tab:key_specifications}}.
    \textbf{(a)}~VSC, including a control PCB and a power PCB, plus one dedicated EMI PCB on the grid side and one on the motor side.
    \textbf{(b)}~CSC, including a control board and a power PCB. The split dc-link inductor can be seen on the left side of the converter, and the converter uses four small overvoltage-protection PCBs.
    Figure adapted from \cite{Neha_CSC_VSC}.}
    \label{fig:hardware}
    \vspace{-1.5em}
\end{figure}

\textbf{Fig.~\ref{fig:hardware}} shows the VSC and CSC hardware demonstrators realized in \cite{Neha_CSC_VSC}, with power stage schematics given in \textbf{Fig.~\ref{fig:schematics}} and specifications and realization details in \textbf{Tab.~\ref{tab:key_specifications}}.
The demonstrators reach volumetric power densities of about \qty{1.7}{kW/dm^3} for the VSC and \qty{1.8}{kW/dm^3} for the CSC, respectively.

The VSC demonstrator shown in \textbf{Fig.~\ref{fig:hardware}a}  is a back-to-back connection of a voltage-source rectifier and a voltage-source inverter sharing a dc-link capacitor and a dc-side common-mode choke.

Two-stage EMI filters on the grid and on load side ensure compliance with  CISPR~11 Class~A and upcoming limits for the lower frequency range of \qty{9}{kHz}\dots\qty{150}{kHz} from IEC~TS~62578~C2.
The realized VSC demonstrator uses commercial \qty{600}{V}, \qty{70}{m\ohm} GaN transistors, a Xilinx Zynq~7000 controller, and a dc link of \qty{500}{V} CeraLink capacitors \cite{Neha_CSC_VSC}. Note that unipolar GaN transistors with the same \qty{140}{m\ohm} rating as the GaN M-BDS used for the CSC (see below) were not available. Thus, the VSC was built with commercial \qty{70}{m\ohm} devices and its semiconductor losses are scaled to \qty{140}{m\ohm} for the comparison; details can be found in \cite{Neha_CSC_VSC}.

\begin{table}[t]
\centering
\caption{Key specifications and realization of the two \qty{1.4}{kW} demonstrators \cite{Neha_CSC_VSC}.}
\label{tab:key_specifications}
\vspace{-1mm}
\begin{tabular}{lrr}
\toprule
Specification & \multicolumn{2}{c}{Common value} \\
\midrule
Rated power & \multicolumn{2}{c}{\qty{1.4}{kW}} \\
Grid / motor voltage & \multicolumn{2}{c}{\qty{200}{V} line-to-line rms} \\
Grid frequency & \multicolumn{2}{c}{\qty{50}{Hz}} \\
Max.\ motor frequency & \multicolumn{2}{c}{\qty{200}{Hz}} \\
Switching frequency & \multicolumn{2}{c}{\qty{72}{kHz}} \\
Nominal ac-ac efficiency & \multicolumn{2}{c}{97\%} \\
Conducted EMI & \multicolumn{2}{c}{CISPR 11 A, IEC TS 62578 C2} \\
\toprule
Realization & VSC & CSC \\
\midrule
Dc-link inductor & --- & $2\times\qty{600}{\micro\henry}$ \\
Dc-link capacitor & \qty{400}{V} & --- \\
Power transistors & GaN, \qty{600}{V} & GaN M-BDS, \qty{600}{V} \\
On-state resistance\textsuperscript{a} & \qty{70}{m\ohm} & \qty{140}{m\ohm} \\
Volumetric power density & \qty{1.7}{kW/dm^3} & \qty{1.8}{kW/dm^3} \\
NEMA PI weighted efficiency\textsuperscript{b} & 95.7\% & 96.6\% \\
\bottomrule
\end{tabular}

\vspace{2mm}\raggedright
{\footnotesize
\textsuperscript{a}The built VSC uses commercial \qty{70}{m\ohm} devices, with losses scaled to \qty{140}{m\ohm} for a comparison at similar chip area~\cite{Neha_CSC_VSC}.\\
\textsuperscript{b}Average over the four operating points from \textbf{Tab.~\ref{tab:nema_points}}, each held for a quarter of the operating time. The VSC value uses semiconductor losses scaled to \qty{140}{m\ohm}~\cite{Neha_CSC_VSC}. The CSC value is for the \qty{140}{m\ohm} M-BDSs as built.}
\end{table}

\begin{table}[t]
\centering
\caption{Operating points of the NEMA Power Index profile for variable-load centrifugal systems. Voltage and current in per unit of the rated motor line-to-line voltage and phase current \cite{nemaPowerIndex2022}.}
\label{tab:nema_points}
\vspace{-1mm}
\begin{tabular}{lcccc}
\toprule
Point & Voltage (p.u.) & Current (p.u.) & Power & Time (p.u.) \\
\midrule
MP1 & 0.44 & 0.19 & \qty{117}{W} & 0.25 \\
MP2 & 0.60 & 0.36 & \qty{302}{W} & 0.25 \\
MP3 & 0.79 & 0.62 & \qty{686}{W} & 0.25 \\
MP4 & 1.00 & 1.00 & \qty{1.4}{kW} & 0.25 \\
\bottomrule
\end{tabular}
\vspace{-1em}
\end{table}

\textbf{Fig.~\ref{fig:hardware}b} shows the CSC demonstrator, which consists of a current-source rectifier on the input side and a current-source inverter on the output side which share a dc-link inductor and a common-mode choke. The dc-link current is set to the larger of the grid-side and the motor-side current amplitudes, so that, advantageously, the dc-link current reduces at part load.
The CSC demonstrator employs first-generation \qty{600}{V}, \qty{140}{m\ohm} GaN M-BDSs.
Each of the four commutation cells, comprising three M-BDSs each, is equipped with an overvoltage-protection PCB, which essentially provides a path for the dc-link current in a fault state where all M-BDSs in a commutation cell would turn-off, e.g., due to loss of gate drive power supply \cite{Neha_CSC_VSC}.

Both demonstrators operate at a switching frequency of \qty{72}{kHz}.
When the semiconductor losses of the VSC are scaled to those expected with \qty{140}{m\ohm} devices, both demonstrators achieve a nominal ac-ac efficiency of 97\%. 
Considering a typical mission profile for variable-load centrifugal systems (as given by the NEMA Power Index \cite{nemaPowerIndex2022}, see \textbf{Tab.~\ref{tab:nema_points}}), the weighted efficiencies are 95.7\% for the VSC and 96.6\% for the CSC \cite{Neha_CSC_VSC}. 
The higher mission-profile efficiency of the CSC can be attributed to lower part-load losses, mainly as a consequence of, first, the ability to reduce the dc-link current with load, and, second, the lower switched voltage of the inverter stage at lower motor voltages.

Finally, note that a comprehensive description of the hardware implementation and experimental characterization of both demonstrators is provided in \cite{Neha_CSC_VSC} and is therefore not repeated here.

\section{Embodied Carbon Footprints}\label{sec:embodied}
Focusing on the most prominent environmental impact indicator, we limit the following discussion to the carbon footprint (or global warming potential, GWP) measured in kilograms of CO\textsubscript{2} equivalents (kg\,CO\textsubscript{2}eq).\footnote{Further environmental impact categories, such as acidification, eutrophication, ozone depletion, and water deprivation, could also be considered in an extended environmental assessment, following frameworks such as the European Union's EF 3.1 Impact Indicators or the International EPD System GPI 5 \cite{EPD_GPI5}.}

\subsection{Component Modeling}
\label{ssec:modeling}
We find the embodied carbon footprint of the two demonstrators based on component-level models for the carbon emissions released to the atmosphere during the material sourcing and manufacturing of a component. The models are based on mass-specific data from the ecoinvent LCA database \cite{ecoinventEcoinventDatabase} and also directly on results reported in the literature. More details are described in \cite{imperialiMultiobjectiveMinimizationLifecycle2024}, and hence, for brevity, we summarize key aspects only. 

Both power stages employ GaN power semiconductors, for which the area-specific embodied GWP is \qty{27000}{kg\,CO_2eq/m^2} \cite{vaucheCradletogateLifeCycle2024}. The corresponding die area of a standard unipolar device is estimated based on its on-state resistance and voltage rating. Note that, although the VSC demonstrator is equipped with \qty{70}{m\ohm} transistors, these are represented in the inventory using the die area corresponding to \qty{140}{m\ohm} devices for consistency with the reported mission-profile efficiencies.
As GaN \mbox{M-BDSs} employ the same manufacturing process as standard GaN transistors, the M-BDS is first modeled assuming equal chip area as a unipolar \qty{140}{m\ohm} GaN transistor, due to the shared drift region.
In the baseline inventory, a unipolar \qty{140}{m\ohm} transistor and an M-BDS are therefore assigned the same embodied carbon footprint.\footnote{An M-BDS can have a somewhat larger die than a unipolar transistor of the same on-state resistance, because of the second gate and the stabilization circuitry. That case is not part of this baseline inventory, but \textbf{Section~\ref{ssec:embodied_carbon_footprint}} shows the sensitivity of the CSC embodied carbon footprint to the assumptions regarding the chip area \cite{Neyer_ECCE_2026,Bisi_GaN_BDS}.}

The GWP contributions of the inductors and cold plate are determined from their respective masses and material compositions, with the inductor cores and windings modeled separately to account for the different materials used in each component. 
The first-stage inductors of the VSC and the dc-link inductor of the CSC are custom assemblies comprising ferrite cores and copper windings, while the common-mode chokes employ nanocrystalline toroidal cores and several of the second-stage inductors are standard catalogue components \cite{Neha_CSC_VSC}.

The capacitors are classified according to their respective technologies: CeraLink ceramic capacitors are used in the VSC dc link, film capacitors in the differential-mode filters, and Class~II MLCCs as common-mode Y-capacitors. Their capacitance and voltage ratings are accounted for indirectly,  through their respective masses.
The PCBs are estimated from the board outline and hence from their area. The finished boards have large cutouts, but the panels are manufactured as solid outlines and the cut-out material is assumed to be waste, therefore the model considers the full outline area and does not credit the cutouts as saved laminate.
The gate drivers, controller, and remaining surface-mounted components are modeled using mass-specific GWP data. 
Further details on the modeling approach are provided in \cite{imperialiMultiobjectiveMinimizationLifecycle2024}.

\subsection{Embodied Carbon Footprints} \label{ssec:embodied_carbon_footprint}

\begin{figure}
    \centering
    \includegraphics[width=\columnwidth]{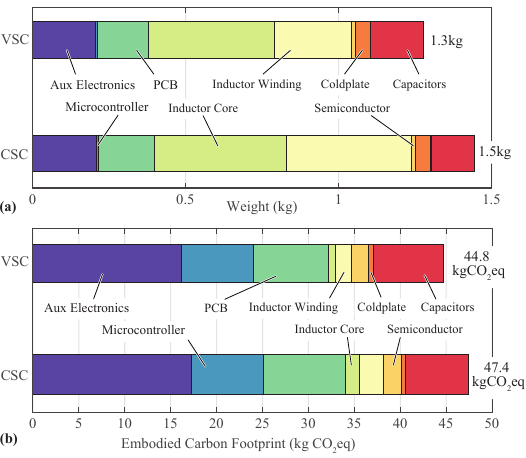}
    \caption{Comparison of \textbf{(a)} mass and \textbf{(b)} embodied carbon footprint breakdowns by component class of the VSC and CSC demonstrators shown in \textbf{Fig.~\ref{fig:hardware}}. 
    }
    \label{fig:weight}
    \vspace{-.5em}
\end{figure}

\begin{figure}
    \centering
    \includegraphics[width=\columnwidth]{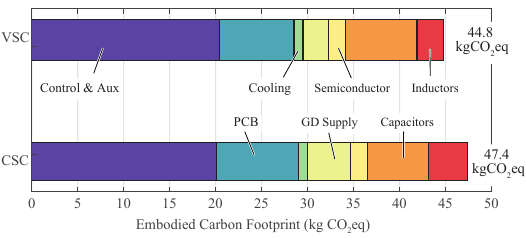}
    \caption{Embodied carbon footprint breakdowns by functional units of the VSC and CSC demonstrators shown in \textbf{Fig.~\ref{fig:hardware}}. Note the increased contributions of the CSC's gate drive power supply as a result of requiring two actively driven gates per M-BDS.
    }
    \label{fig:functional_groups}
    \vspace{-1em}
\end{figure}

\textbf{Fig.~\ref{fig:weight}a} compares the mass distribution of the two demonstrators. With a total mass of \qty{1.5}{kg}, the CSC is slightly heavier than the VSC (\qty{1.3}{kg}), primarily due to its greater amount of magnetic components. The contributions of the PCBs and capacitors are approximately equal for both converters. In both demonstrators, the twelve GaN transistors and the control electronics account for only a small fraction of the total mass. 

\textbf{Fig.~\ref{fig:weight}b} reports the corresponding embodied carbon footprints of the two demonstrators.
The totals lie close together, i.e., \qty{44.8}{kg\,CO_2eq} for the VSC and \qty{47.4}{kg\,CO_2eq} for the CSC respectively.
Although the magnetic components dominate the mass, their contribution to the embodied carbon footprint is much less pronounced, in line with results from similar studies \cite{Imperiali_dc_dc}.
Auxiliary electronics and the microcontroller emerge as the main contributors, which is attributed to the energy-intensive manufacturing processes of the integrated circuits.
Similarly the relatively large PCBs add a further substantial share.

A different breakdown of the carbon footprint according to functional groups is provided in \textbf{Fig.~\ref{fig:functional_groups}}. 
As discussed previously, the M-BDSs are modeled using the die area of an equivalent unipolar \qty{140}{m\ohm} device and since both converters employ twelve switches, their semiconductor contributions are identical. 
The CSC, however, requires a more complex gate-drive arrangement, as each M-BDS has two independent gates, which require dedicated gate signals and gate drive circuitry. 
Both demonstrators employ the same AMD Zynq~7000 system on chip (SoC), consequently, the increased control complexity does not result in a larger contribution of the SoC to the carbon footprint.
Instead, the additional hardware requirements are reflected in increased contributions of the gate drivers and the gate-drive supplies. 
The functional breakdown further shows that control and auxiliary electronics contribute the largest share in both converters, while the CSC exhibits a larger contribution from magnetic components.

In an M-BDS, both voltage polarities share a common drift region within a single die, suggesting that its area can be comparable to that of a unipolar device with the same on-state resistance. Hence, so far, the semiconductor contribution to the GWP is based on the first-order assumption that the M-BDS and an equivalent (i.e., same on-state resistance) unipolar GaN transistor have the same chip area. 
However, the additional gate and substrate stabilization circuitry \cite{Neyer_ECCE_2026} may increase the chip area beyond this ideal estimate and since quantitative data regarding this increase is limited, a sensitivity analysis is provided here. 
For example, a recent study suggests that the chip area of a GaN M-BDS may be up to 1.8 times that of a unipolar device with the same on-state resistance \cite{Bisi_GaN_BDS}. 
To account for this uncertainty, \textbf{Fig.~\ref{fig:mbds_footprint}a} compares the embodied carbon footprint of the semiconductors and the required gate drivers only, first for the ideal case (assuming equal chip areas), and for a second case where the M-BDS chip area corresponds to 1.8 times that of the unipolar devices of the same on-state resistance used in the VSC. Clearly, the CSC shows a higher embodied carbon footprint for its power transistors including gate drive circuitry in both cases. Note, however, that the worst-case increase of around \qty{1.5}{kg\,CO_2eq} for a 1.8 times higher chip area of the M-BDS would not strongly impact the system-level results reported in \textbf{Fig.~\ref{fig:weight}b} and \textbf{Fig.~\ref{fig:functional_groups}}.

\subsection{Self-Reverse-Blocking M-BDS}
\label{ssec:self-reverse-blocking}

\begin{figure}
    \centering
    \includegraphics[width=\columnwidth]{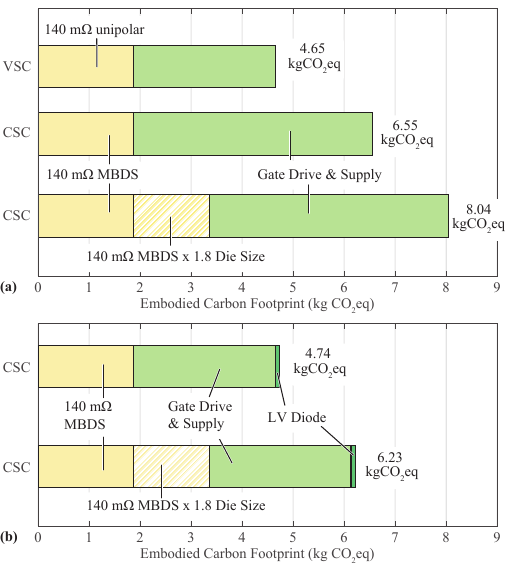}
    \caption{Embodied carbon footprints of the GaN power transistors and their gate-drive circuitry used in the VSC and the CSC.
    \textbf{(a)}~Sensitivity on M-BDS chip area increase compared to a unipolar GaN transistor of same on-state resistance (\qty{140}{m\ohm} baseline and chip area ratios 1:1 and 1:1.8). 
    \textbf{(b)} Impact of employing self-reverse-blocking M-BDS (SRB-M-BDS) in the CSC, where only one of the two M-BDS gates requires a gate drive circuitry while the other is controlled using a low-voltage Schottky cascode diode (see  \textbf{Fig.~\ref{fig:mbds}b}).
    }
    \label{fig:mbds_footprint}
    \vspace{-1em}
\end{figure}

\begin{figure}[t]
    \centering
    \includegraphics[width=\columnwidth]{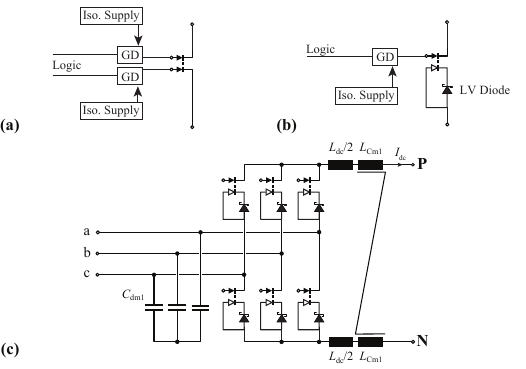}
    \caption{Gate-drive arrangements for an M-BDS.
    \textbf{(a)}~Both gates are of the normally-off type and driven actively, i.e., require dedicated gate drive circuitry (isolated power supply and gate drive IC, and dedicated logic signals).
    \textbf{(b)}~Self-reverse-blocking M-BDS \cite{Nain_Self_Reverse_Blocking_Monolithic_Bidirectional_GaN_Transistor} with one normally-on gate automatically controlled via a low-voltage diode in cascode configuration, realizing the behavioral equivalent of a unipolar transistor with a series diode, however with lower losses. 
    \textbf{(c)}~CSC rectifier realized with self-reverse-blocking M-BDSs and unidirectional dc-link current.}
    \label{fig:mbds}
    \vspace{-1em}
\end{figure}

\textbf{Fig.~\ref{fig:mbds_footprint}} clearly shows the impact  of the additional gate drivers and isolated power supplies required to drive the CSC's M-BDSs, which increase both circuit complexity and embodied carbon footprint.
Actively controlling both gates of the M-BDSs as shown in \textbf{Fig.~\ref{fig:mbds}a} comes further at the price of requiring dedicated multi-step commutation sequences \cite{Neha_CSC_VSC}, but gives full flexibility in terms of control, and in particular allows operation with both dc-link current directions.
Whereas the inverse dc-link current direction is necessary in, e.g., dc-ac current-source inverters for realizing bidirectional power flow, an ac-ac CSC can realize both power flow directions for a given, constant direction of the dc-link current by swapping the polarity of the the dc-side \emph{voltages}.
Hence, the switching elements only need to provide unidirectional current conduction but still bipolar voltage blocking capability. 

This motivates the consideration of alternative gate-drive configurations for the M-BDS in ac-ac CSCs, specifically the self-reverse-blocking M-BDS concept illustrated in \textbf{Fig.~\ref{fig:mbds}b} \cite{Nain_Self_Reverse_Blocking_Monolithic_Bidirectional_GaN_Transistor}. 
One of the M-BDS gates is realized as a normally-on gate (with correspondingly lower, negative threshold voltage), and hence can be controlled via a low-voltage Schottky diode in cascode configuration to mimic diode operation. Advantageously, the low-voltage diode only marginally increases the on-state voltage drop of the device, while only one gate drive circuitry and only one logic signal are required; further there is no need for multi-step commutation sequences anymore.

A resulting CSC rectifier implementation is shown in \textbf{Fig.~\ref{fig:mbds}c}, where the number of gate signals and isolated supplies is reduced by half compared to the baseline in \textbf{Fig.~\ref{fig:schematics}}, matching the gate drive effort of the VSC.
The corresponding impact on the embodied carbon footprint is evaluated in \textbf{Fig.~\ref{fig:mbds_footprint}b}. 
While the additional low-voltage diodes contribute only marginally to the overall carbon footprint, the gate-drive effort is almost cut in half.

\section{Use Phase}\label{sec:use_phase}
As discussed in \textbf{Section~\ref{sec:embodied}}, the difference in the embodied carbon footprint of the two demonstrators is only \qty{2.6}{kg\,CO_2eq}. 
Over the lifetime of a motor drive, the energy losses associated with power conversion can result in substantially higher emissions, as the energy dissipated within the converter is drawn from the electricity grid and therefore carries the corresponding grid carbon footprint. 

Therefore, we evaluate the life-cycle carbon footprints of the two demonstrators based on the use-phase carbon emissions under both, a carbon-intensive electricity mix and a renewable electricity mix. Since the considered VSDs are intended to supply motors driving variable-load centrifugal systems, which typically operate predominantly at partial load, the VSD losses should be evaluated over a representative mission profile rather than solely at rated power; we thus  consider the  NEMA PI mission profile  for variable-load centrifugal systems \cite{nemaPowerIndex2022}, see \textbf{Tab.~\ref{tab:nema_points}}. 
Based on these operating points, a weighted efficiency can be defined as the ratio of the weighted output power to the weighted input power \cite{huberLifecycleCarbonFootprints2024} as
\begin{equation}
\eta_{\mathrm{w}}
=
\frac{\sum_{i} w_i P_{i}}{\sum_{i} w_i \left(P_{\mathrm{loss},i}+P_{i}\right)},
\label{eq:eta_w}
\end{equation}
where $w_i$ denotes the fraction of the total operating time spent at operating point $i$, $P_{i}$ is the output power delivered to the load, and $P_{\mathrm{loss},i}$ represents the corresponding converter losses. The weighted efficiency therefore accounts for the different power levels and their respective contributions to the overall energy consumption.

The total energy loss dissipated over a mission profile can subsequently be calculated from the efficiency at each operating point according to
\begin{equation}
W_\mathrm{loss} = \sum_{i=1}^{N} \Delta t_i\frac{1-\eta_i}{\eta_i}\, P_i,
\label{eq:loss}
\end{equation}
where $\eta_i$ is the converter efficiency at operating point $i$, and $\Delta t_i$ is the time spent in operating point $i$. 
For the NEMA PI mission profile $w_i = 1/4~\forall~i$, we have $\Delta t_i=T_\mathrm{mission}/4$ with $T_{\mathrm{mission}}$ denoting the total operating time of the converter. Accordingly, using $\eta_{\mathrm{NEMA}}$ to denote the weighted efficiency $\eta_{\mathrm{w}}$ of this profile, \eqref{eq:loss} can be expressed as
\begin{equation}
W_\mathrm{loss} = \frac{T_\mathrm{mission}}{4}\,\frac{1-\eta_{\mathrm{NEMA}}}{\eta_{\mathrm{NEMA}}} \sum_{i=1}^{4}  P_i.
\label{eq:loss_nema}
\end{equation}

Both demonstrators are assumed to operate continuously, i.e.,  24~hours per day and 7 days per week, over a period of $T_{\mathrm{mission}} = \qty{10}{years}$. Furthermore, the conversion efficiencies are assumed to remain unchanged throughout the operational lifetime (i.e., no performance degradation is considered). The resulting use-phase carbon footprint is obtained by multiplying the total dissipated energy by the carbon intensity of the available grid electricity mix, $\sigma_\mathrm{el}$:
\begin{equation}
\mathrm{GWP}_\mathrm{loss} = W_\mathrm{loss} \, \sigma_\mathrm{el}.
\label{eq:gwp}
\end{equation}

As discussed in \textbf{Section~\ref{sec:hwimplementation}}, both converters achieve similar efficiencies at rated power, whereas the CSC exhibits higher efficiency under partial-load operation. This difference is reflected in their weighted efficiencies under the NEMA PI mission profile, which are \qty{95.7}{\percent} for the VSC and \qty{96.6}{\percent} for the CSC, as reported in \textbf{Tab.~\ref{tab:key_specifications}}. 

\begin{figure}
    \centering
    \includegraphics[width=\columnwidth]{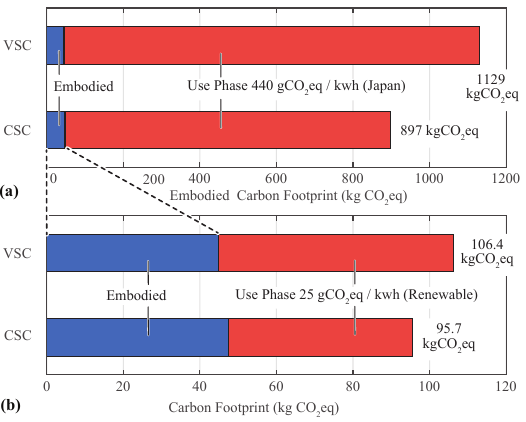}
    \caption{Life-cycle carbon footprints of the VSC and CSC demonstrators for \qty{10}{years} of continuous operation according to the NEMA PI mission profile for variable-load centrifugal systems. 
    \textbf{(a)} Current Japanese electricity mix with a carbon intensity of \qty{440}{g\,CO_2eq/kWh} \cite{ieaJapanElectricity}.
    \textbf{(b)} Future renewable electricity mix with a carbon intensity of \qty{25}{g\,CO_2eq/kWh} \cite{Energyscope_Renewable_Emissions}, resulting in significantly lower life-cycle carbon footprints and a much higher importance of the embodied contribution.}
    \label{fig:usephase}
    \vspace{-1em}
\end{figure}

Accordingly, \textbf{Fig.~\ref{fig:usephase}} combines the embodied and use-phase carbon footprints for both converters and for the two considered grid electricity mixes. 
For the current Japanese grid electricity mix with a carbon intensity of \qty{440}{g\,CO_2eq/kWh} \cite{ieaJapanElectricity}, the use-phase contribution dominates the overall life-cycle carbon footprints of both converters, resulting in around \qty{1130}{kg\,CO_2eq} for the VSC and around \qty{900}{kg\,CO_2eq} for the CSC. 
The higher partial-load efficiency of the CSC therefore translates into a substantially lower life-cycle carbon footprint under a carbon-intensive grid, despite its initially slightly higher embodied carbon footprint.

In contrast, when considering a future renewable electricity mix with only \qty{25}{g\,CO_2eq/kWh}, a value that can be reached by combining hydro, wind, and photovoltaic generation \cite{Energyscope_Renewable_Emissions}, the use-phase contribution is considerably reduced, making the embodied carbon footprint a comparable share of the overall impact.
Specifically, the resulting life-cycle footprints decrease to \qty{106}{kg\,CO_2eq} for the VSC and \qty{96}{kg\,CO_2eq} for the CSC. The difference between the two converters becomes considerably smaller, as the benefits of the higher partial-load efficiency of the CSC are reduced when the carbon intensity of the electricity supply is low.

\section{Conclusion}\label{sec:conclusion}

Monolithic bidirectional GaN power transistors enable ac-ac CSCs for VSD applications which are competitive with state-of-the-art VSC realizations. This paper complements prior research comparing efficiency, power density, EMI compliance, and functionality of ac-ac VSC and CSC demonstrators (1.4\,kW, 200\,V line-to-line rms), including grid-side and motor-side EMI filters, with a comparative life-cycle assessment. 

For both demonstrators, the life-cycle carbon footprint is primarily determined by the use-phase emissions for a grid electricity mix with high carbon intensity. Whereas reducing mission-profile losses thus remains the main design objective for continuous operation under these conditions, a future decarbonized electricity supply leads to comparable relative importance of the embodied and use-phase emissions. Therefore, identifying the main contributors to the embodied carbon footprint is important for assessing potential design improvements and simplifications, such as the reduction of the gate drive circuitry of the CSC by applying the self-reverse-blocking M-BDS concept.

Overall, the results motivate an application-specific evaluation of converter topologies, considering the expected mission profile, operating lifetime, and electricity supply already during the design process. Further improvements should target not only semiconductor performance but also the realization effort of the surrounding circuitry, which contributes substantially to the embodied carbon footprint of both topologies.